\documentstyle[epsf,aps,prl,preprint]{revtex}

\begin{document}

\draft
\author{S.-C.~Gou, J.~Steinbach and P.L.~Knight}
\address{Optics Section, Blackett Laboratory, Imperial College, 
London SW7 2BZ, United Kingdom}
\title{Dark pair coherent states of the motion of a trapped ion}
\date{\today}
\maketitle
\begin{abstract}
We propose a scheme for generating vibrational pair coherent states of
the motion of an ion in a two-dimensional trap. In our scheme, the trapped 
ion is excited bichromatically by three laser beams along different 
directions in the $X$-$Y$ plane of the ion trap. 
We show that if the initial vibrational state is given by a two-mode
Fock state, the final steady state, indicated by the extinction 
of the fluorescence emitted by the ion, is a pure state. The motional
state of the ion in the equilibrium realizes that of the 
highly-correlated pair coherent state.
\end{abstract}
\pacs{42.50.Vk, 42.50.Dv, 32.80.Pj}

A variety of generalized coherent states has been constructed to describe different 
physical phenomena \cite{coherent}. Mathematically, the constructions of
these sets of generalized coherent states are associated with particular Lie
groups. The Glauber coherent states, defined as the right
eigenstates of a single-mode boson annihilation operator are associated with
the Heisenberg group. Beyond these canonical coherent states, particular
generalized coherent states associated with the noncompact SU(1,1) group have been 
extensively studied \cite{coherent,agarwal-1,barut,perelomov,horn,SU11}. According to 
Barut and Girardello \cite{barut}, these associated coherent states are defined 
as the eigenstates of the SU(1,1) lowering operator, whereas in the definition 
given by Perelomov \cite{perelomov}, they are generated by the SU(1,1) analogue 
of the displacement operator. These two sets of coherent states are different 
though they are closely related to the SU(1,1) group.

Regarding the two-mode boson realization of the SU(1,1) group, a special set
of coherent states of the Barut-Girardello type known as pair coherent
states (PCS) can be formulated \cite{coherent}. If $\hat{a}$ 
($ \hat{a}^\dagger $) and $\hat{b}$ ($\hat{b}^\dagger$) denote two
independent boson annihilation (creation) operators, then $\hat{a}\hat{b}$ 
($\hat{a}^{\dagger }\hat{b}^{\dagger}$) stands for the pair annihilation
(creation) operator for the two modes. The pair coherent states
$| \xi ,q\rangle_{\text{PCS}}$ are defined as eigenstates of both the pair 
annihilation operator $\hat{a}\hat{b}$ and the number difference operator 
$\hat{Q}=\hat{a}^{\dagger }\hat{a}-\hat{b}^{\dagger }\hat{b}$, i.e., 
\begin{equation}
\hat{a}\hat{b}|\xi ,q\rangle_{\text{PCS}} =\xi |\xi ,q\rangle_{\text{PCS}} 
\text{ , }\hat{Q} |\xi ,q\rangle_{\text{PCS}} =q|\xi ,q\rangle_{\text{PCS}}\,,
\end{equation}
where $\xi $ is a complex number and $q$ is the ``charge'' parameter which
is a fixed integer. Furthermore, the PCS can be expanded as a superposition 
of the two-mode Fock states,
\begin{equation}
|\xi ,q\rangle_{\text{PCS}} ={\cal N}_q\sum_{l=0}^\infty \frac{\xi ^l}
{\sqrt{l!\left( l+q\right) !}}|l+q,l\rangle _{\text{F}}\,,
\end{equation}
where ${\cal N}_q=[ \left| \xi \right| ^{-q}I_q\left( 2\left| \xi
\right| \right) ]^{-1/2}$ is the normalization constant ($I_q$ is
the modified Bessel function of the first kind of order $q$). Pair coherent
states were introduced by Horn and Silver \cite{horn} to describe the
production of pions and applied to other problems in quantum field theory 
\cite{coherent}. In quantum optics, PCS are regarded as an important type of 
correlated two-mode state, which possess prominent nonclassical properties 
such as sub-Poissonian statistics, correlation in the number fluctuations,
squeezing, and violations of Cauchy-Schwartz inequalities \cite{agarwal-1}.

The experimental realization of such nonclassical states is of practical
importance. Agarwal \cite{agarwal-1} suggested that the optical PCS can be 
generated via the competition of 4-wave mixing and two-photon absorption in
a nonlinear medium. This is the only scheme proposed to generate PCS known
to us. In this Rapid Communication we propose a scheme to generate 
vibrational PCS of a trapped ion. Recently, due to remarkable advances in 
laser cooling of a trapped ion \cite{Diedrich}, it has become possible to
realize nonclassical states of the centre-of-mass (CM) motion of a single 
trapped ion. An ion confined in an electromagnetic trap can be regarded as 
a particle with quantized CM motion moving in a harmonic potential. Exciting 
or deexciting the internal atomic states of the trapped ion by a classical 
laser driving field changes the external states of the ion motion, as atomic 
stimulated absorption and emission processes are always accompanied by 
momentum exchange of the laser field with the ion.  If both the vibrational
amplitude of the ion is much smaller than the laser wavelength, i.e., in 
the Lamb-Dicke limit \cite{Diedrich}, and the driving field is tuned to one
of the vibrational sidebands of the atomic transition, then this model can
be simplified to a form similar to the Jaynes-Cummings model (JCM)
\cite{Blockley} in which the quantized radiation field is replaced by the
quantized CM motion of the ion. As the coupling between the vibrational
modes and the external environment is extremely weak, dissipative effects
which are inevitable from cavity damping in the optical regime, can be
significantly suppressed for the ion motion. This unique feature thus
makes it possible to realize cavity QED experiments without using an optical
cavity. Following this approach, nonclassical vibrational states of the
trapped ions such as Fock \cite{number}, squeezed \cite{squeeze} and
Schr\"{o}dinger cat states \cite{even-odd,poyatos} have been proposed and
observed \cite{meekhof}.

\begin{figure}
   \caption[]
   {
   \leavevmode
   \epsfxsize=140mm
   \epsffile{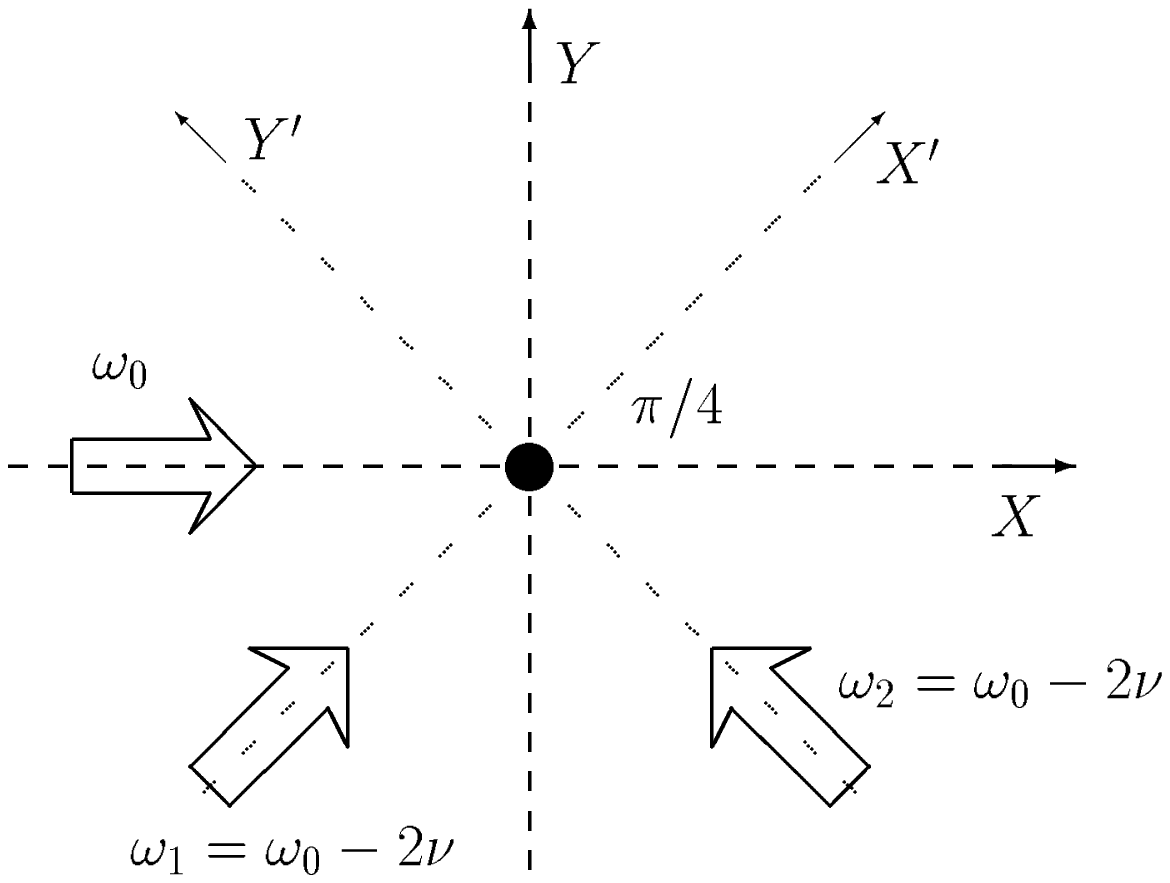}  
   }
   { 
	Configuration to generate vibrational PCS. The carrier
	field drives the ion on resonance along the $X$-axis. The other two
    lasers are both resonant with the second lower vibrational sideband.
	They drive the ion along the $X^\prime$ and $Y^\prime$-axis having 
	phases $\phi_1=0$ and $\phi_2=\pi$ respectively.
   }
   \label{fig1}
\end{figure}

Consider the quantized motion of a two-level ion of mass $M$ which is trapped
in a two-dimensional (2-D) isotropic harmonic potential characterized by the
trap frequency $\nu $. The creation (annihilation) of vibrational quanta in
the $X$ and $Y$ directions is described by the operators
$\hat{a}\ (\hat{a}^{\dagger })$ and $\hat{b}\ (\hat{b}^{\dagger })$
respectively. The position operators are given by 
$\hat{x} =\lambda (\hat{a}+\hat{a}^{\dagger }) $and $\ \hat{y}=\lambda
(\hat{b}+\hat{b}^{\dagger }) $, where $\lambda =\sqrt{\hbar /2\nu M}.$ In our
scheme, which we sketch in Fig.\ref{fig1}, the ion is driven bichromatically
by three laser beams in the $X$-$Y$ plane. The first two lasers are both
tuned to the second lower vibrational sideband and applied to the ion along
directions with an angle $\pi /4$ and 3$\pi /4$ relative to the $X$-axis,
respectively. The third laser which drives the ion along the $X$-axis is 
resonant with the atomic transition frequency. In the rotating-wave
approximation, the Hamiltonian describing the coherent evolution is 
\begin{equation}
\hat{H}=\hbar \nu \left( \hat{a}^{\dagger }\hat{a}+\hat{b}^{\dagger }\hat{b}%
\right) +\frac{\hbar \omega _0}2\hat{\sigma}_z-\left[ {\cal D}E^{(-)}
\left( \hat{x},\hat{y},t\right) \hat{\sigma}_{-}+\text{ h.c.}%
\right]\,,
\end{equation}
where the first two terms describe the free evolution of the external and
internal degrees of freedom of the ion and the last indicates the atom-field
interaction. The operators $\hat{\sigma}_{+}$ and $\hat{\sigma}_{-}$ are
raising and lowering operators for the two-level ion obeying commutation
relations $\left[ \hat{\sigma}_{+},\hat{\sigma}_{-}\right] =\hat{\sigma}_z$ and
$\left[ \hat{\sigma}_z,\hat{\sigma}_{\pm }\right] =\pm 2\hat{\sigma}_{\pm }.$
The transition in the two-level ion is characterized by the dipole matrix
element ${\cal D}$ and the transition frequency $\omega _0.$ The negative
frequency part of the classical electric driving field is given by 
\begin{eqnarray}
E^{\left( -\right) }\left( \hat{x},\hat{y},t\right) &=&E_1e^{i\left[ \left(
\omega _0-2\nu \right) t-k_2\hat{x}^{\prime }+\phi _1\right]
}+E_2e^{i\left[ \left( \omega _0-2\nu \right) t-k_2\hat{y}^{\prime }+\phi
_2\right] }  \nonumber \\
&&+E_0e^{i\left( \omega _0t-k_0\hat{x}+\phi _0\right)}\,,
\end{eqnarray}
where $2 E_j$ and $\phi _j$ indicate the amplitudes and phases of the driving
lasers respectively. We have introduced new position operators 
$ \hat{x}^{\prime }$ and $\hat{y}^{\prime }$ which are related to $\hat{x}$
and $\hat{y}$ by a $\pi /4$ rotation in the $X$-$Y$ plane, 
\begin{equation}
\left( 
\begin{array}{c}
\hat{x}^{\prime } \\ 
\hat{y}^{\prime }
\end{array}
\right) =\frac 1{\sqrt{2}}\left( 
\begin{array}{cc}
1 & 1 \\ 
-1 & 1
\end{array}
\right) \left( 
\begin{array}{c}
\hat{x} \\ 
\hat{y}
\end{array}
\right)\,,
\end{equation}
so that for the creation (annihilation) operators $\hat{A}$ 
$(\hat{A}^{\dagger }) \;$and $\hat{B}$ $( \hat{B}^{\dagger })\,,$ defined in
the $X^{\prime }$ and $Y^{\prime }$ directions respectively, we obtain the 
following transformation 
\begin{equation}
\left( 
\begin{array}{c}
\hat{A} \\ 
\hat{B}
\end{array}
\right) =\frac 1{\sqrt{2}}\left( 
\begin{array}{cc}
1 & 1 \\ 
-1 & 1
\end{array}
\right) \left( 
\begin{array}{c}
\hat{a} \\ 
\hat{b}
\end{array}
\right)\,.
\end{equation}

According to Vogel and de Matos Filho \cite{even-odd,vogel2}, if the ion is 
in the resolved sideband limit and the driving laser is resonant with one of 
the vibrational sidebands, then the ion-laser interaction can be described as 
a nonlinear JCM. Following Refs.\ \cite{even-odd,vogel2}, the Hamiltonian of 
Eq.(3) can be written in the interaction picture as 
\begin{eqnarray}
\hat{H}_{\text{I}} &=&\hbar\,e^{-\eta ^2/2}\left[ \sum_{j=0}^\infty \frac{\left(
i\eta \right) ^{2j+2}}{j!(j+2)!}\left[\Omega _1e^{i\phi _1} \hat{A}^j
(\hat{A}^\dagger)^{j+2}+\Omega _2e^{i\phi _2} \hat{B}^j (\hat{B}^\dagger)^{j+2}\right] 
\right.  \nonumber \\
&&\left. +\Omega _0e^{i\phi _0}\sum_{j=0}^\infty \frac{\left( i\eta \right)
^{2j}}{j!j!} \hat{a}^j (\hat{a}^\dagger)^j\right] \hat{\sigma}%
_{-}+\text{ h.c.}\,,
\end{eqnarray}
where $\Omega _j=-{\cal D} E_j/\hbar $ are the Rabi frequencies, and the
Lamb-Dicke parameter $\eta =k\lambda $ has been defined assuming $k_2\simeq
k_0=k.$

As the damping of vibrational quanta can be significantly suppressed in an 
ion trap, the dominant decay process is the spontaneous emission from the 
two-level ion, and the time evolution of the system in the interaction
picture can be described by a density operator $\hat{\rho}$ obeying the 
master equation \cite{number,even-odd} 
\begin{equation}
\frac{d\hat{\rho}}{dt}=-\frac i\hbar \left[ \hat{H}_{\text{I}},\hat{\rho}%
\right] +\frac \Gamma 2\left( 2\hat{\sigma}_{-}\hat{\varrho}\hat{\sigma}_{+}-%
\hat{\sigma}_{+}\hat{\sigma}_{-}\hat{\rho}-\hat{\sigma}_{-}\hat{\sigma}_{+}%
\hat{\rho}\right)\,,
\end{equation}
where $\Gamma $ is the spontaneous decay rate of the excited state of the ion, 
and the modified density operator 
\begin{equation}
\hat{\varrho}=\frac 14\int_{-1}^1\int_{-1}^1dudvW(u,v)e^{ik\left( u\hat{x}+v%
\hat{y}\right) }\hat{\rho}e^{-ik\left( u\hat{x}+v\hat{y}\right)}\,,
\end{equation}
accounts for the momentum transfer in the $X$-$Y$ plane due to spontaneous
emission, where $W\left( u,v\right) $ describes the angular distribution of
the spontaneous emission. In the Lamb-Dicke regime, $\eta \ll 1$, the
master equation, Eq.(8), can be well approximated by the first order
expansion in $\eta .$ In this case, only the leading terms, i.e., $j=0$ in $%
\hat{H}_{\text{I}}$, are considered, and $\hat{\varrho}$ is replaced by $%
\hat{\rho}.$ Setting $\Omega _1=\Omega _2=\Omega $, and $\phi _1=0 ,\phi
_2=\pi$, the master equation Eq.(8) simplifies to 
\begin{equation}
\frac{d\hat{\rho}}{dt}=-\frac i\hbar \left[ \hat{H}_{\text{I}}^{\prime },%
\hat{\rho}\right] +\frac \Gamma 2\left( 2\hat{\sigma}_{-}\hat{\rho}\hat{%
\sigma}_{+}-\hat{\sigma}_{+}\hat{\sigma}_{-}\hat{\rho}-\hat{\sigma}_{-}\hat{%
\sigma}_{+}\hat{\rho}\right)\,,
\end{equation}
with the effective Hamiltonian $\hat{H}_{\text{I}}^{\prime }$ given by 
\begin{equation}
\hat{H}_{\text{I}}^{\prime }=\alpha \left[ \hat{a}\hat{b}-\xi \right] \hat{%
\sigma}_{+}+\text{ h.c.}\,,
\end{equation}
where $\alpha =\hbar \Omega \eta ^2\exp \left( -\eta ^2/2\right) $, and $\xi
=\Omega _0\Omega ^{-1}\eta ^{-2}\exp \left(- i\phi _0\right) .$
Disregarding the constant driving term $\xi $, the effective Hamiltonian of
Eq.(11) is identical to the non-degenerate two-mode JCM \cite{gou1}.

For a master equation of the form of Eq.(10), the steady-state solution 
$\hat{\rho}_{\text{s}}$ is a pure state \cite{number,squeeze,even-odd}
\begin{equation}
\hat{\rho}_{\text{s}}=|g\rangle |\psi \rangle \langle \psi|
\langle g|\,,
\end{equation}
where $|g\rangle $ is the atomic ground state and $|\psi \rangle $ defines the
vibrational state of the ion. When the system reaches the steady state
$d\hat{\rho}_{\text{s}}/dt=0$. Assuming a steady state of the form above
gives $[ \hat{H}_{\text{I}}^{\prime },\hat{\rho}_{\text{s}}] =0,$ as the
dissipative term on the right hand side of Eq.(10) vanishes in this case.
Consequently, a sufficient condition for us to generate the vibrational steady
state $|\psi \rangle $ is to ensure $\hat{\rho}_{\text{s}}$ satisfies the
restriction $[ \hat{H}_{ \text{I}}^{\prime },\hat{\rho}_{\text{s}}] =0,$ 
or equivalently $ \hat{a}\hat{b}|\psi \rangle =\xi |\psi \rangle\,.$
Evidently, this is identical to our first definition of a PCS given above
in Eq.(1). In order to generate a fully characterized PCS, however, we need
another constraint to ensure that the final vibrational state remains an
eigenstate of the number difference operator $\hat{Q}$ [Eq.(1)]. This can
be achieved by properly choosing the initial vibrational state. In view of
the fact that $[ \hat{a}\hat{b},\hat{Q}] =0,$ it follows that the number
difference is a constant of motion in processes involving simultaneous pair
annihilation or creation. Thus for an initial vibrational state
$| \psi_0 \rangle$, if the conserved ``charge'' condition is initially
satisfied, i.e., $\hat{Q}| \psi_0 \rangle =q| \psi_0 \rangle $, then this
condition holds for the whole time evolution of the system described by Eq.(10). 
Regarding the feasibility of an experimental realization, it seems that the
two-mode Fock state $| \psi_0 \rangle =| m+q,m\rangle _{\text{F}}$ is best
suited to fulfil the conserved ``charge'' condition. Without loss of
generality we may set the initial vibrational state as $| \psi _0\rangle
=|q,0\rangle _{\text{F}}$ ($q \ge 0$) which, according to  recent work, can be
prepared with very high efficiency \cite{meekhof}.

In order to identify the validity of our analytic argument and to gain 
insight into the transient behaviour we have solved the master equation,
Eq.(10), numerically employing a Monte-Carlo state-vector method
\cite{monte-carlo}. Our numerical analysis was performed using a high-order 
unravelling technique \cite{steinbach95} in a finite (truncated) Fock state
basis with a cut-off chosen such that an increase of this cut-off does not
alter the result of our integration. 

\begin{figure}
   \caption[]
   { 
   \leavevmode
   \epsfxsize=140mm
   \epsffile{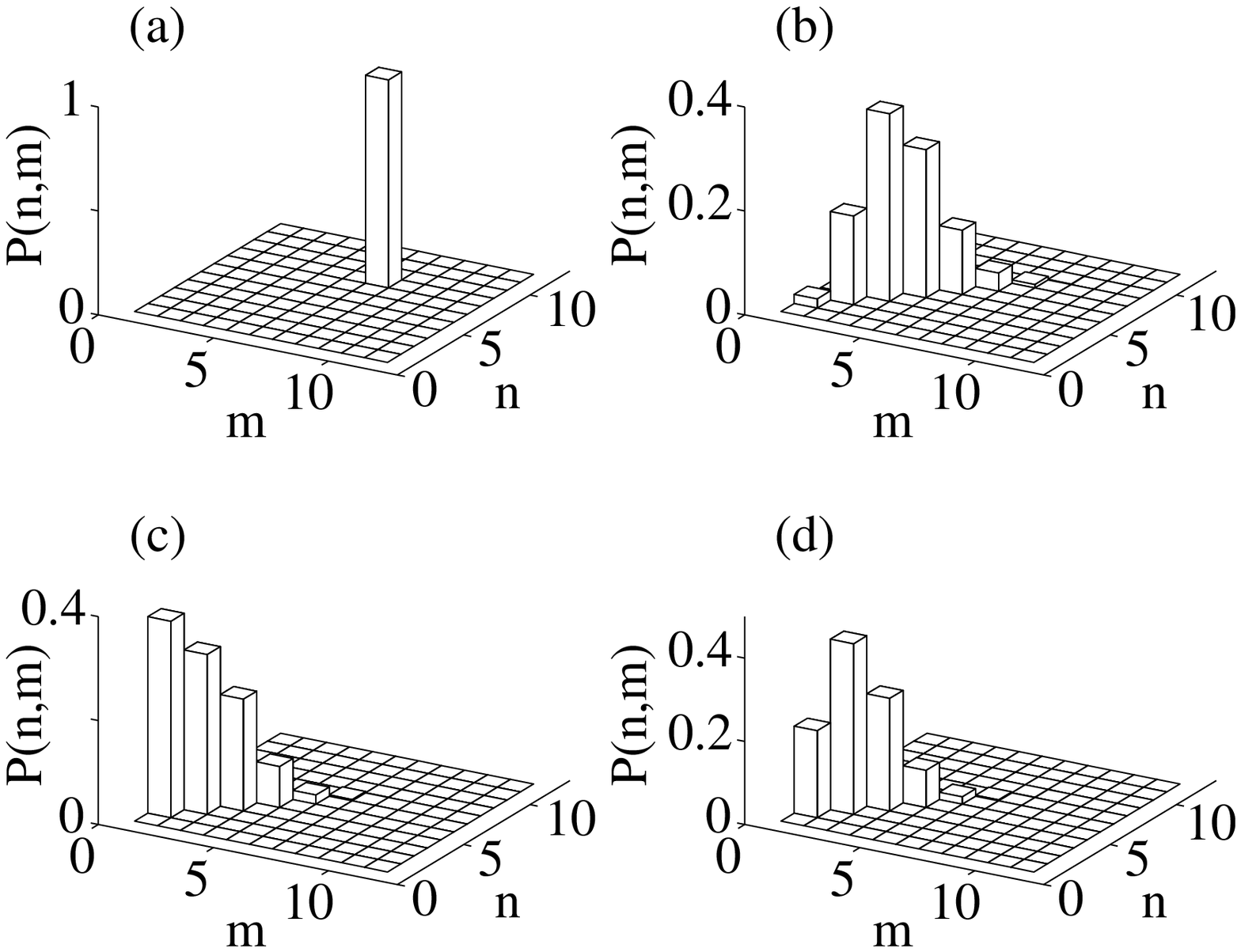}
   }
   { 
	We depict the excitation number distribution P(n,m) in the two vibrational
	modes at different times. The system evolves from (a), the initial pure state
	$|\Psi_0\rangle = |e\rangle \otimes |7,6\rangle_{\text{F}}$ ($\Gamma t = 0.0$),
	through (b) and (c), two intermediate states ($\Gamma t = 125.0$) and 
	($\Gamma t = 500.0$), into (d), its steady state which is the pure product
	of the internal ground state of the ion and a vibrational PCS: $|\Psi_{\text{s}}
	\rangle = |g\rangle \otimes |2,1\rangle_{\text{PCS}}$ ($\Gamma t = 2000.0$).
	The data shown has been obtained from a Monte-Carlo simulation which included
	1000 trajectories and was performed in a truncated Fock state basis 
	($|0,0\rangle_{\text{F}} ... |20,20\rangle_{\text{F}}$).
	Parameters: $\alpha = 0.2, \xi = 2.0, \Gamma = 10.0\,.$
   }
   \label{fig2}
\end{figure}

\begin{figure}
   \caption[]
   { 
   \leavevmode
   \epsfxsize=140mm
   \epsffile{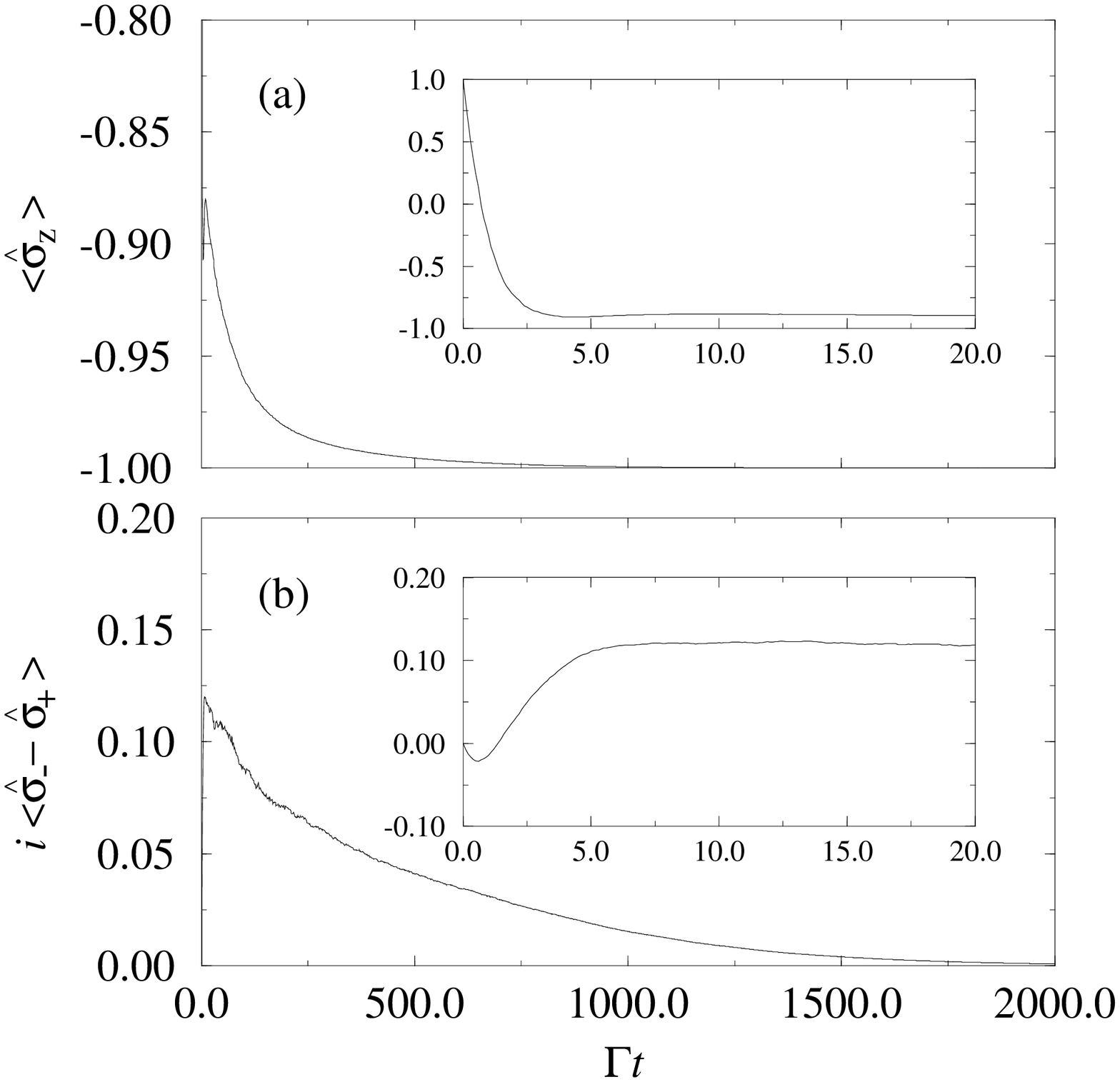}
   }
   {
	Time evolution of the internal state of the two-level ion as the system 
	evolves into its steady state. (a) shows the inversion, 
	$\langle \hat{\sigma}_z \rangle,$ and (b) the imaginary part of the polarization, 
	$i\langle \hat{\sigma}_{\text{-}}-\hat{\sigma}_{\text{+}}\rangle\,.$ 
	The real part of the polarization, 
	$\langle \hat{\sigma}_{\text{-}}+\hat{\sigma}_{\text{+}} \rangle,$
	remains zero at all times. The two insets show the evolution
	in the time interval $\Gamma t = 0.0$ to $\Gamma t = 20.0\,.$ Parameters
	and data used are the same as in Fig.\ref{fig2}\,.
   }
   \label{fig3}
\end{figure}

We find that the system evolves from an initial product of a two-mode Fock
state and a superposition of the internal states of the ion into the pure steady
state [Eq.(12)] as expected. Fig.\ref{fig2} depicts the excitation number
distribution in the two vibrational modes at different times, showing the 
system as it evolves from the initial pure state $|\Psi _0\rangle = |e\rangle
\otimes |7,6\rangle _{\text{F}}$ into its steady state. The steady-state 
excitation number distribution [Fig.\ref{fig2}(d)] is indistinguishable from 
that of a PCS with $\xi=2$ and $q=1.$ This is the expected result since the
particular choice of the initial state determines the conserved ``charge''
$q = \langle \Psi_0| \hat{Q}|\Psi _0\rangle =1\,,$ and from Eqs.(1) and (11)
the second number defining the PCS is determined by the driving term $\xi$ in
the effective Hamiltonian, which in the example shown is $\xi = 2\,.$ We verify
that the steady state is pure by calculating the trace of the square of the 
density operator Tr$\left( \hat{\rho}^2_{\text{s}}\right) $ from the steady
state obtained in our numerical integration. We find Tr$\left( \hat{%
\rho}^2_{\text{s}}\right) = 0.9997\,.$ Fig.\ref{fig3} shows the transient 
behaviour of the internal state of the ion. We depict the time evolution of 
the inversion, $\langle \hat{\sigma}_z\rangle,$ [Fig.\ref{fig3}(a)] and the 
imaginary part of the polarization, $i\left\langle \hat{\sigma}_{-}-\hat{\sigma}_{+}
\right\rangle,$ [Fig.\ref{fig3}(b)]. We note the timescale for the system 
to reach the steady state is much longer than it would take the same ion to
reach its steady state in a resonance fluorescence experiment.

One may examine the existence of the PCS by observing the collapses and the
revivals of the atomic inversion. This is illustrated by switching off the
carrier field suddenly once the system has reached the steady state. In this
case, the system is equivalent to the non-degenerate two-mode JCM
interacting with a PCS when the ion is initially in its ground state. The
time evolution of such a system has been investigated by several authors
\cite{gou2}. In particular, when $q=0$ the Rabi oscillation of the atom exhibits
exactly periodic behaviour. However, as the atomic decay is essential to the
present treatment and cannot be ignored, the collapses and revivals of the
Rabi oscillation of the trapped ion are substantially different from the
previous results. One possible way to suppress the complications caused by the
atomic decay is to increase the intensities of the driving lasers so that
the ratio $\Gamma /\alpha $ is lowered. Thus, in the short time regime the
influence of spontaneous emission can be eliminated.

In conclusion, we have proposed a scheme for the realization of pair
coherent states of the CM motion of a trapped ion. In our scheme, three
laser beams, one of which is tuned to the carrier frequency and the other
two to the second lower vibrational sideband, are used to drive the ion
trapped in a 2-D isotropic harmonic potential well. In appropriate limits,
the system will relax to a steady state due to the spontaneous emission from
the ion. If the vibrational state of motion of the ion is initially
prepared in a Fock state, then the steady state of the system is a pure
state given by a product of the atomic ground state with a PCS of the
vibrational motion. In this case, the two parameters, $\xi $ and $q$, that
characterize the PCS are determined by the intensities and phases of the
driving lasers and by the number difference between the two vibrational
modes determined by the initial Fock state respectively.

This work was supported in part by the UK Engineering and Physical Sciences
Research Council and the European Union. S.-C. Gou acknowledges support
from the Ministry of Education, Taiwan, Republic of China. J. Steinbach is
supported by the German Academic Exchange Service (DAAD-Doktorandenstipendium
aus Mitteln des zweiten Hochschulsonderprogramms).


\end{document}